\documentclass[reprint,amsmath,amssymb,aps,prl,floatfix,superscriptaddress,nofootinbib]{revtex4-1}

\usepackage{graphicx}
\usepackage{booktabs}
\usepackage{multirow}
\usepackage{dcolumn}
\usepackage{makecell}
\usepackage{mathtools}
\usepackage{tabularx}
\usepackage{colortbl}
\usepackage{physics}

\usepackage[colorlinks,linkcolor=blue,anchorcolor=blue,citecolor=blue, urlcolor=blue]{hyperref}

\usepackage{multirow}

\begin{document}

\title{Tightening constraints on primordial oscillations with latest ACT and SPT data}

\author{Ze-Yu Peng}
\email{pengzeyu23@mails.ucas.ac.cn}
\affiliation{School of Physical Sciences, University of Chinese Academy of Sciences, Beijing 100049, China}
\affiliation{International Centre for Theoretical Physics Asia-Pacific, University of Chinese Academy of Sciences, 100190 Beijing, China}

\author{Yun-Song Piao}
\email{yspiao@ucas.ac.cn}
\affiliation{School of Physical Sciences, University of Chinese Academy of Sciences, Beijing 100049, China}
\affiliation{International Centre for Theoretical Physics Asia-Pacific, University of Chinese Academy of Sciences, 100190 Beijing, China}
\affiliation{School of Fundamental Physics and Mathematical Sciences, Hangzhou Institute for Advanced Study, UCAS, Hangzhou 310024, China}
\affiliation{Institute of Theoretical Physics, Chinese Academy of Sciences, P.O. Box 2735, Beijing 100190, China}

\begin{abstract}

The oscillation feature in primordial power spectrum (PPS), a fingerprint of not only a wide class of models
of inflation but new physics, 
is of significant theoretical interest, and can be imprinted on
the cosmic microwave background (CMB). In this work, we present
constraints on periodic oscillations in the PPS using the
latest ACT DR6 and SPT-3G D1 CMB data with the precise
measurements at high multipoles beyond the Planck angular
resolution and sensitivity. It is found that the combination of
SPT and ACT with Planck CMB dataset significantly tightens the
upper bound to $A_\mathrm{log,lin}\lesssim 0.029$ at $95\%$ C.L., showing
no hint for primordial oscillations, where $A_\mathrm{log,lin}$ are the
amplitudes of logarithmic and linear oscillation in the PPS,
respectively. Our work presents state-of-the-art CMB constraints
on primordial oscillations, highlighting the power of the
ground-based CMB experiments in constraining physics beyond the
simplest slow-roll models.

\end{abstract}

\maketitle

\newpage

\textit{Introduction.---}
Inflation
\cite{Guth:1980zm,Linde:1981mu,Albrecht:1982wi,Starobinsky:1980te},
the standard paradigm of the very early universe, predicts a
nearly scale-invariant primordial power spectrum (PPS) of scalar
perturbations. Recent Planck cosmic microwave background (CMB)
data \cite{Planck:2018vyg} strongly support the simplest slow-roll
inflationary scenarios, and its combination with recent BICEP/Keck
data \cite{BICEP:2021xfz} has placed stringent constraints on the
landscape of models \cite{Planck:2018jri}.

However, physics beyond simplest slow-roll potentials
\cite{Starobinsky:1992ts,Adams:2001vc,Wang:2002hf,Gong:2005jr,Slosar:2019gvt,Achucarro:2022qrl}
can result in the oscillations in the PPS,
which also are consistent with current data, see also models with
null energy condition violation
\cite{Cai:2020qpu,Cai:2022nqv,Cai:2023uhc}. In particular, the
periodic oscillation feature is of significant theoretical
interest, motivated by such as Trans-Planckian effects
\cite{Martin:2000xs,Easther:2002xe,Danielsson:2002kx,Baumann:2014nda},
axion monodromy inflation in string theory
\cite{Silverstein:2008sg,McAllister:2008hb,Flauger:2009ab}, and
primordial standard clocks
\cite{Chen:2011zf,Chen:2015lza,Chen:2018cgg}.
There have been some earlier searches for such oscillations with
CMB data
\cite{Easther:2004vq,Meerburg:2013cla,Meerburg:2013dla,Peiris:2013opa,Easther:2013kla,Aich:2011qv}.
Currently, the most stringent constraints on primordial
oscillations were reported by the Planck cooperation based on
their CMB data \cite{Planck:2018jri}, $A_\mathrm{log,lin}\lesssim 0.038$
at $95\%$ C.L., where $A_\mathrm{log,lin}$ are the amplitudes of
logarithmic and linear oscillation in the PPS respectively, which
recently has been complemented by constraints from large-scale
structure (LSS) surveys,
e.g.\cite{Beutler:2019ojk,Ballardini:2022wzu,Mergulhao:2023ukp}.
See also \cite{Braglia:2021ckn,Braglia:2021sun,Braglia:2021rej,Braglia:2022ftm,Petretti:2024mjy} for constraints on other realistic models of primordial features. 

Recently, both Atacama Cosmology Telescope (ACT)
\cite{ACT:2025fju, ACT:2025tim} and South Pole Telescope (SPT)
\cite{SPT-3G:2025bzu}, the ground-based CMB experiments, have
released their new data, offering the most precise measurements of
small-scale polarization spectra to date. Their combination with
Planck data yields the tightest CMB constraints, which shows no
evidence for physics beyond $\Lambda$CDM. These high-precision
high-$\ell$ polarization data are expected to be powerful for
constraining primordial oscillations, given their lower foreground
contamination and sharper transfer functions compared to the
temperature spectrum \cite{Chluba:2015bqa}, see
Ref.~\cite{Antony:2024vrx} for a search for primordial
oscillations using the previous SPT-3G data.

\begin{figure*}[htbp]
    \centering
   \includegraphics[width=0.45\linewidth]{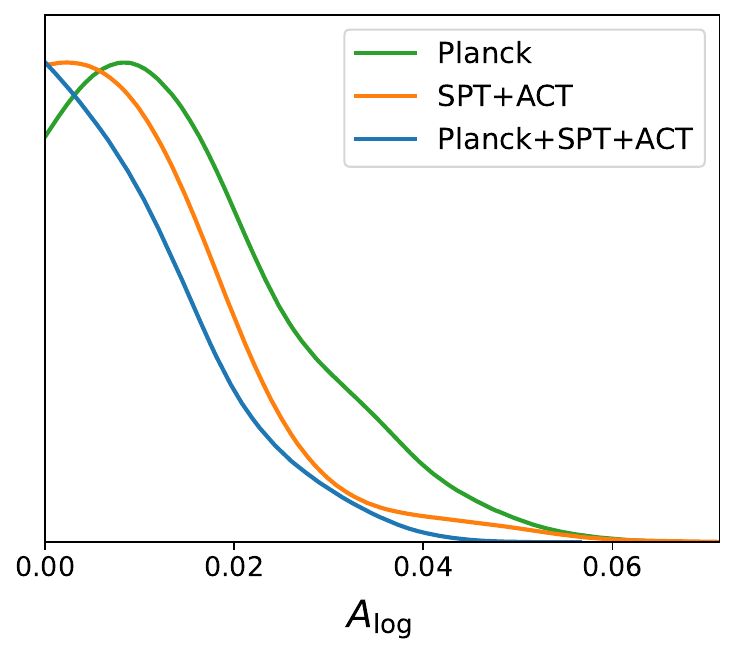}
   \includegraphics[width=0.45\linewidth]{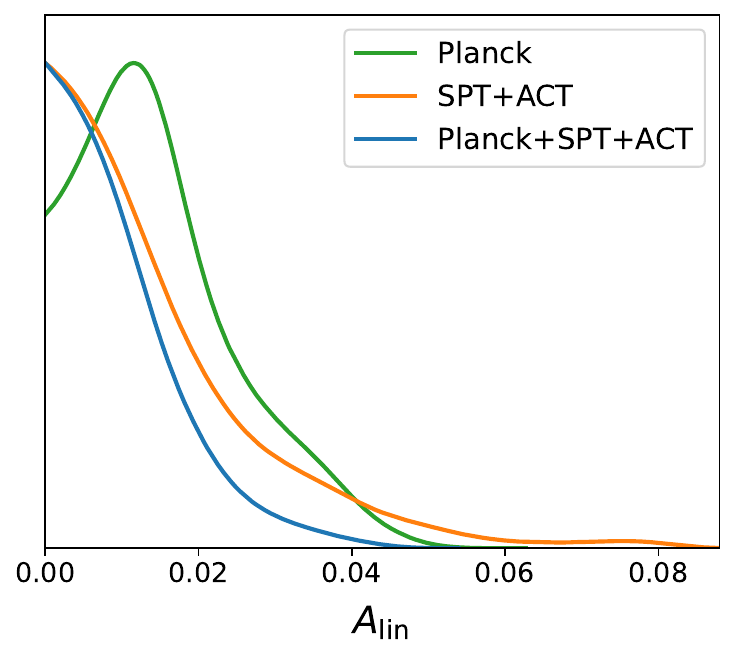}
\caption{1D posterior distributions of the logarithmic (left) and
linear (right) oscillating amplitudes, obtained using Planck,
SPT+ACT, and Planck+SPT+ACT datasets, respectively. }
    \label{fig:A_X}
\end{figure*}

In this work, we present state-of-the-art constraints on
primordial oscillations leveraging the latest ACT DR6 and SPT-3G
D1 data. As shown in Fig.~\ref{fig:A_X}, we find that the
constraints set by SPT+ACT on primordial
oscillations are comparable to those from Planck, and the
combined Planck+SPT+ACT dataset significantly tightens the
upper bound to $A_\mathrm{log,lin}\lesssim 0.029$ at $95\%$ C.L., showing
no hint for primordial oscillations.


\textit{Primordial oscillations.---}
Typically, different new physics beyond the simplest slow-roll
models can make the PPS manifest different oscillatory patterns.
The sudden changes in the inflaton potential, such as bumps and
steps
\cite{Starobinsky:1992ts,Adams:2001vc,Gong:2005jr,Hazra:2010ve} or
sharp turns
\cite{Ashoorioon:2008qr,Achucarro:2010da,Konieczka:2014zja}, can
imprint linear oscillations onto the PPS, while the resonant
models \cite{Chen:2008wn} and
the axion monodromy model
\cite{Silverstein:2008sg,McAllister:2008hb,Flauger:2009ab} can
lead to logarithmically oscillating signals in the PPS.

\begin{figure}[htbp]
    \centering
   \includegraphics[width=0.45\textwidth]{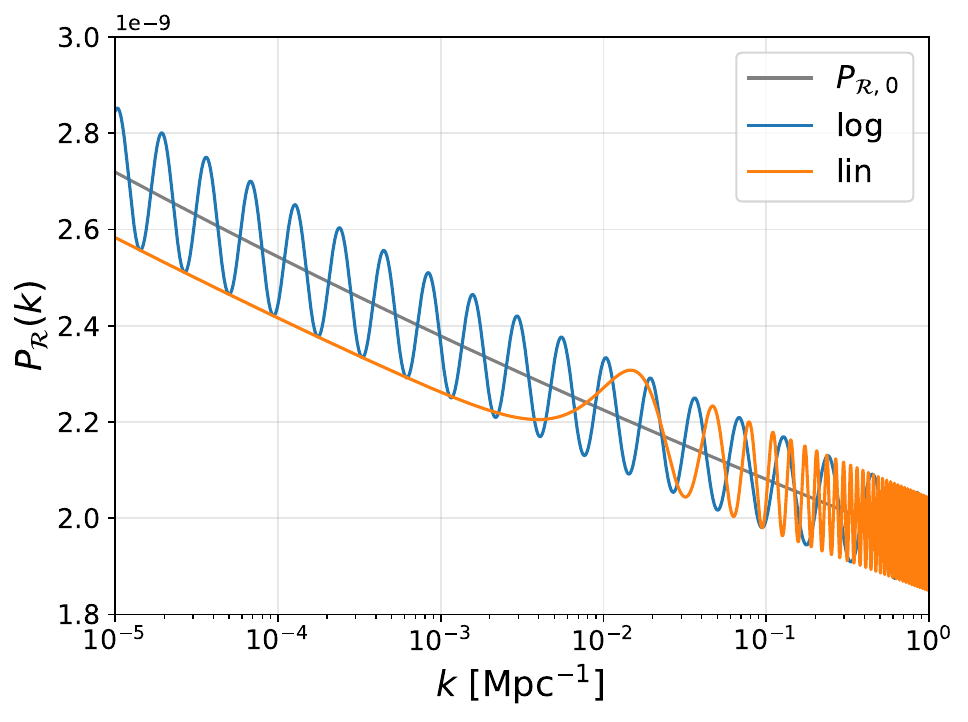}
\caption{Primordial power spectra with the logarithmic and linear
oscillations, where $A_{lin,log}=0.05$, $\omega_{lin,log}=10$, and
$\phi_{lin,log}=3.14$. }
    \label{fig:feature}
\end{figure}

The templates we consider for the corresponding linear and
logarithmical oscillating PPS are,
e.g.\cite{Planck:2018jri,Slosar:2019gvt}, as shown in
Fig.~\ref{fig:feature}:
\begin{equation}
P_{\mathcal{R}}(k) = P_{\mathcal{R},0}(k)\left[1 + A_l
\cos\left(\omega_l({\frac{k}{k_*}})_l +\phi_l\right)\right],
\end{equation}
where $P_{\mathcal{R},0}(k) =
A_\mathrm{s}(k/k_*)^{n_\mathrm{s}-1}$ is the standard power-law
PPS, the pivot scale is $k_* = 0.05\,\mathrm{Mpc}^{-1}$, and
$l=\mathrm{log}$ and $({k/k_*})_{\mathrm{log}} = \log(k/k_*)$ for
logarithmic oscillation, while $l=\mathrm{lin}$ and $({k/k_*})_{\mathrm{lin}} = k/k_*$ for linear oscillation.


\textit{Datasets and Methods.---}
In our analysis, inspired by \cite{SPT-3G:2025bzu}, we consider
three datasets as follows:
\begin{itemize}
\item \textbf{Planck}: The \texttt{Plik-lite} likelihood for
Planck PR3 high-$\ell$ temperature and polarization data, and the
\texttt{Commander} likelihood for low-$\ell$ TT and the
\texttt{SimALL} likelihood for low-$\ell$ EE
\cite{Planck:2019nip}, as well as the Planck PR4 CMB lensing data
\cite{Carron:2022eyg}. \item \textbf{SPT+ACT}: The
foreground-marginalized
\texttt{SPT-lite}\footnote{\url{https://github.com/SouthPoleTelescope/spt_candl_data}}
likelihood for SPT-3G D1 \cite{SPT-3G:2025bzu,Balkenhol:2024sbv}, combined with the
\texttt{ACT-lite}\footnote{\url{https://github.com/ACTCollaboration/DR6-ACT-lite}}
likelihood for ACT DR6 \cite{ACT:2025fju, ACT:2025tim}, as well as
the ACT DR6 CMB lensing data\footnote{We use the
\texttt{actspt3g\_baseline} variant of
\url{https://github.com/qujia7/spt_act_likelihood}.}
    \cite{ACT:2023kun,ACT:2023dou,ACT:2023ubw} and SPT-3G \cite{SPT-3G:2024atg,SPT-3G:2025zuh}. Here, we impose a Gaussian prior on the optical depth, $\tau_\mathrm{reio} = 0.054 \pm 0.007$ \cite{Planck:2018vyg}.
\item \textbf{Planck+SPT+ACT}: The combination of Planck and
SPT+ACT, as well as the CMB lensing data\footnote{We use the
\texttt{actplanckspt3g\_baseline} variant of
\url{https://github.com/qujia7/spt_act_likelihood}}. Following 
Refs.\cite{ACT:2025fju, SPT-3G:2025bzu}, we cut the Planck CMB
data to $\ell_{max}=1000$ for TT and $\ell_{max}=600$ for TE and
EE.
\end{itemize}
In addition, we include the DESI DR2 baryon acoustic oscillations
(BAO) dataset \cite{DESI:2025zgx} and the Pantheon+ supernova (SN)
dataset \cite{Scolnic:2021amr}.

\begin{table}[htbp]
\centering
\begin{tabular}{|l|c|}
\hline
Parameter  & Prior \\
         \hline
$A_X$ & $[0, 0.5]$ \\
$\omega_X$ & $[0, 100]$ \\
$\phi_X$ & $[0, 2\pi]$ \\
\hline
$\log(10^{10} A_\mathrm{s})$ & $[1.61, 3.91]$ \\
$n_\mathrm{s}$ & $[0.8, 1.2]$ \\
$H_0$ & $[20, 100]$ \\
$\Omega_\mathrm{b} h^2$ & $[0.005, 0.1]$ \\
$\Omega_\mathrm{c} h^2$ & $[0.001, 0.99]$ \\
$\tau_\mathrm{reio}$ & $[0.01, 0.8]$ \\
\hline
\end{tabular}
\caption{The priors for relevant parameters in our MCMC analysis.
Note that a Gaussian prior, $\tau_\mathrm{reio} = 0.054 \pm
0.007$, is applied only to the SPT+ACT dataset.}
\label{tab:priors}
\end{table}

We perform the Markov chain Monte Carlo (MCMC) analysis using
\texttt{Cobaya} \cite{Torrado:2020dgo}. The observables are
computed using the cosmological Boltzmann code \texttt{CLASS}
\cite{Blas:2011rf}. The Gelman-Rubin criterion
\cite{Gelman:1992zz} for all chains is converged to $R-1<0.02$.
The flat priors adopted for relevant parameters are presented in
Table \ref{tab:priors}. The best-fit parameters and corresponding
$\chi^2$ values are obtained using the global optimization method
in \texttt{PROSPECT} \cite{Holm:2023uwa}.


\textit{Results.---}
In Table~\ref{tab:AX}, we see that compared to the Planck results,
SPT+ACT yields a slightly tighter constraint on the logarithmic
oscillating amplitude $A_\mathrm{log}$, but a higher upper limit
on $A_\mathrm{lin}$ due to the long posterior tail. The combined
Planck+SPT+ACT dataset yields our most stringent constraints,
$A_\mathrm{log} < 0.0286$ and $A_\mathrm{lin} < 0.0267$ ($95\%$
C.L.).

\begin{table}[htbp]
\centering
\begin{tabular}{|l|c|c|c|}
\hline
Dataset & Planck & SPT+ACT & Planck+SPT+ACT \\
\hline
$A_\mathrm{log}$ & $< 0.0387$ & $< 0.0335$ & $< 0.0286$ \\
$A_\mathrm{lin}$ & $< 0.0357$ & $< 0.0433$ & $< 0.0267$ \\
\hline
\end{tabular}
\caption{The $95\%$ upper limits on the oscillating amplitudes
from different datasets.} \label{tab:AX}
\end{table}


\begin{figure}[htbp]
    \centering
   \includegraphics[width=0.45\textwidth]{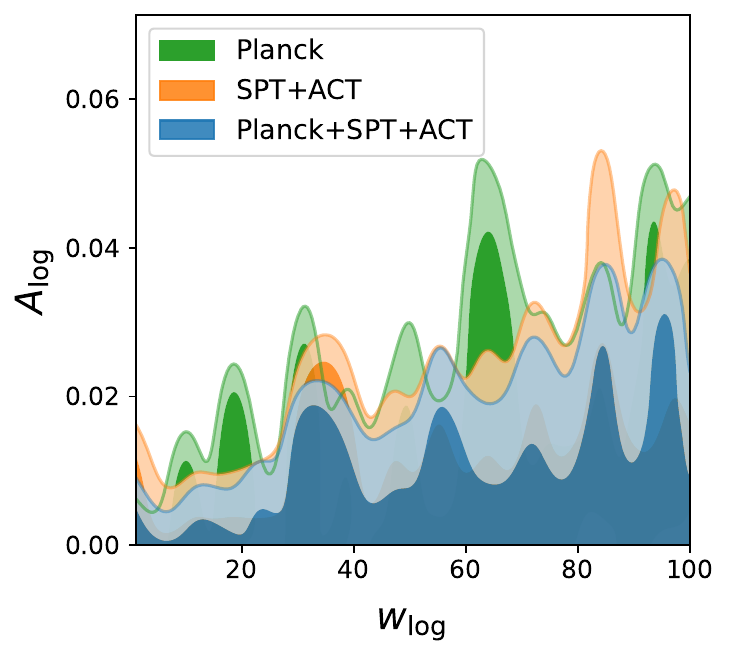}
   \includegraphics[width=0.45\textwidth]{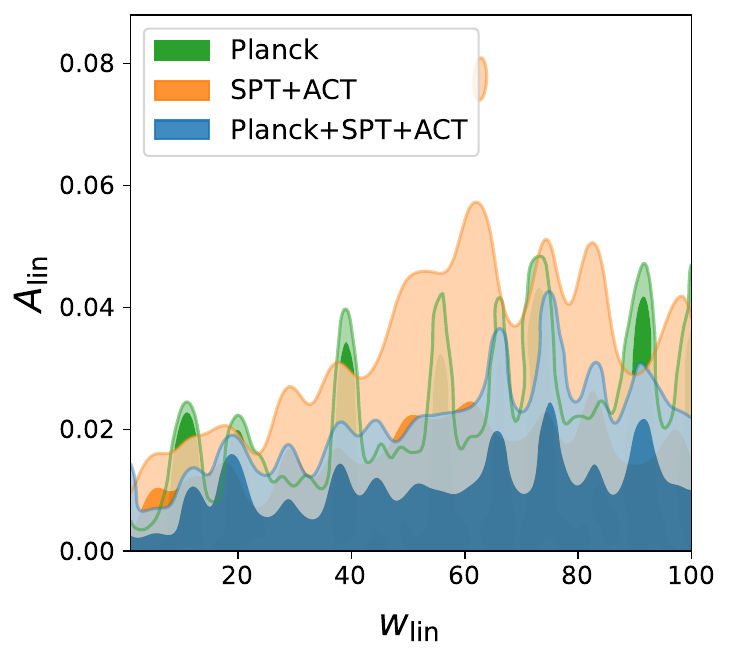}
\caption{The 2D posterior distribution of $A_l$-$\omega_l$ ($68\%$
and $95\%$ confidence levels) for logarithmic (upper) and linear
(lower) oscillating spectra, obtained using Planck, SPT+ACT, and
Planck+SPT+ACT datasets, respectively.}
    \label{fig:AX-wX}
\end{figure}

In Fig.~\ref{fig:AX-wX}, we show the posterior distributions of
$A_\mathrm{log,lin}$-$\omega_\mathrm{log,lin}$. Some
localized $A_\mathrm{log,lin}$ peaks can be observed  at specific frequencies, but they do
not match between the results of Planck and SPT+ACT. This suggests
that these apparent peaks are more likely statistical fluctuations
rather than real physical signals. The combined Planck+SPT+ACT
dataset tightens the constraints across all frequencies, with the
$2\sigma$ upper limits consistently below $0.05$.

\begin{table*}[htbp]
\centering
\begin{tabular}{|l|cc|cc|cc|}
\hline
\multicolumn{1}{|l|}{\textbf{Dataset}} & \multicolumn{2}{c|}{\textbf{Planck}} & \multicolumn{2}{c|}{\textbf{SPT+ACT}} & \multicolumn{2}{c|}{\textbf{\begin{tabular}{@{}c@{}}Planck+\\SPT+ACT\end{tabular}}} \\
\cline{2-7}
 & \textbf{log} & \textbf{lin} & \textbf{log} & \textbf{lin} & \textbf{log} & \textbf{lin} \\
\hline
low-$\ell$ TT & $-0.43$ & $-0.33$  & --- & --- & $0.07$ & $-0.48$ \\
low-$\ell$ EE & $0.41$ & $0.00$ & --- & --- & $-0.27$ & $-1.32$ \\
Planck high-$\ell$ & $-8.04$ & $-10.33$ & --- & --- & $-2.74$ & $-2.20$ \\
SPT-3G D1 & --- & --- & $-3.73$ & $0.08$ &$-0.57$ & $-1.25$ \\
ACT DR6 & --- & --- & $-2.68$ & $-6.79$ & $-4.84$ & $-1.24$ \\
CMB lensing & $0.51$ & $-0.00$ & $-0.19$ & $-0.11$ & $0.04$ & $-0.39$ \\
DESI BAO & $-1.08$ & $0.33$ & $-0.32$ & $0.26$ & $0.13$ & $0.56$ \\
Pantheon+ SN & $0.38$ & $0.09$ & $-0.04$ & $0.06$ & $-0.01$ & $-0.11$ \\
\hline
\textbf{Total $\Delta \chi^2$}    & $\mathbf{-8.26}$ & $\mathbf{-10.97}$ & $\mathbf{-6.11}$ & $\mathbf{-7.16}$ & $\mathbf{-8.33}$ & $\mathbf{-6.43}$ \\
\textbf{Preference level} & $\mathbf{1.7\sigma}$ & $\mathbf{2.3\sigma}$ & $\mathbf{1.3\sigma}$ & $\mathbf{1.5\sigma}$ & $\mathbf{1.8\sigma}$ & $\mathbf{1.3\sigma}$ \\
\hline
\end{tabular}
\caption{$\Delta\chi^2$ values for the best-fit models with
logarithmic and linear oscillations, relative to models with the
standard PPS, obtained using the Planck, SPT+ACT, and
Planck+SPT+ACT datasets, respectively. Note that the CMB lensing data
used for each dataset is different. } \label{tab:chi2}
\end{table*}

To evaluate the statistical significance for primordial
oscillations, we calculate the $\Delta \chi^2$ values for the
best-fit models of the oscillating PPS relative to the standard
PPS, as summarized in Table~\ref{tab:chi2}. The Planck data
indicate some hints for primordial oscillations, we have
$\Delta\chi^2_{\mathrm{log}}=-8.26$ and
$\Delta\chi^2_{\mathrm{lin}}=-10.97$ for logarithmic and linear
oscillations, respectively. This corresponds that primordial
oscillations are preferred at the $1.7\sigma$ and $2.3\sigma$
significance levels. However, for SPT+ACT, the corresponding
values are suppressed to $\Delta\chi^2_{\mathrm{log}} = -6.11$
($1.3\sigma$) and $\Delta\chi^2_{\mathrm{lin}} = -7.16$
($1.5\sigma$). The further inclusion of Planck data into SPT+ACT
does not lead to any significant $\Delta \chi^2$ shift,
$\Delta\chi^2_{\mathrm{log}} = -8.33$ ($1.8\sigma$) and
$\Delta\chi^2_{\mathrm{lin}} = -6.43$ ($1.3\sigma$).
Therefore, essentially latest SPT+ACT dataset suppress the
oscillations in the PPS hinted by Planck, so no hints for
primordial oscillations from the Planck+SPT+ACT datasets.

\begin{figure*}[htbp]
    \centering
   \includegraphics[width=\linewidth]{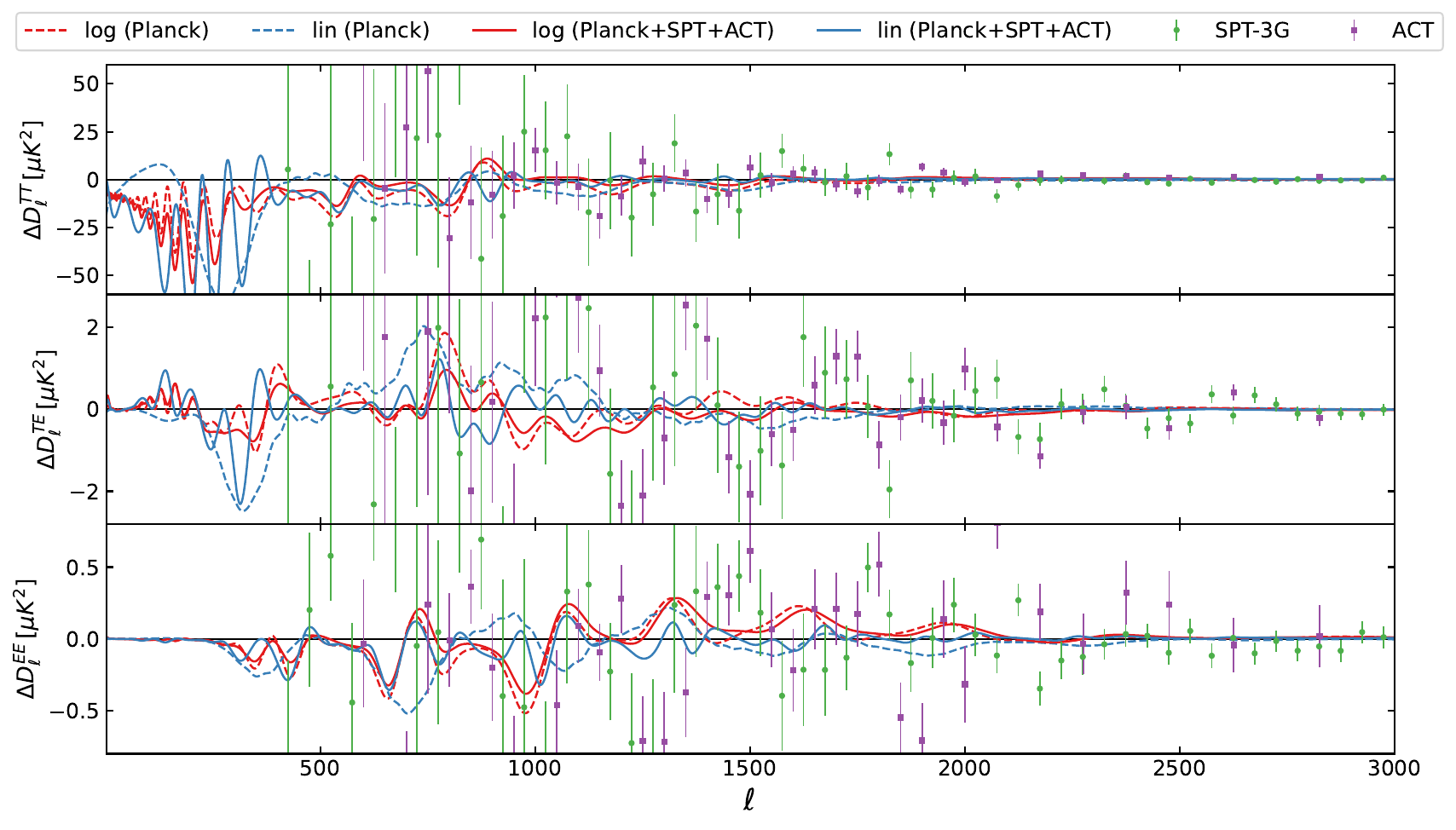}
\caption{The CMB spectral residuals of the best-fit models with
logarithmic and linear oscillations obtained using Planck (dashed)
and Planck+SPT+ACT (solid), respectively. The base line is the
Planck best-fit model without oscillations. The ACT and SPT-3G
data points are also showed.}
    \label{fig:spectrum}
\end{figure*}

In Fig.~\ref{fig:spectrum}, we show the spectral residuals of the
best-fit models with oscillations in the PPS for Planck (dashed)
and Planck+SPT+ACT (solid) datasets respectively, relative to the
Planck best-fit model with standard power-law spectrum. As
expected, the impact of the primordial oscillations is prominent
in the small-scale polarization spectra, and there is a slight
suppression of the residuals around $l \sim 500-1000$ in the TE
and EE spectra for Planck+SPT+ACT compared to Planck results.

\textit{Conclusion.---}
The search for primordial oscillations is of significant
theoretical interest and would open a crucial observational window
into the physics of inflation beyond the simplest slow-roll
models. In this work, we have reported the constraints on
primordial oscillations with the latest ACT DR6 and SPT-3G D1
datasets. The combination of SPT and ACT with Planck dataset yield
our most stringent constraints, the upper limits on the
oscillating amplitudes of $A_\mathrm{log} < 0.0286$ and
$A_\mathrm{lin} < 0.0267$ ($95\%$ C.L.).

Our work presents state-of-the-art CMB constraints on primordial
oscillations, highlighting the power of the ground-based CMB
experiments, especially when combined with large-scale Planck
data. It is expected that upcoming CMB observations, such as the
Simons Observatory \cite{SimonsObservatory:2018koc} and CMB-S4
\cite{Abazajian:2019eic} will further improve our constraints on
primordial oscillations and bring deeper insights into the physics
of inflation. It is well known that LSS also offer a vital,
independent probe to primordial oscillations, especially those
from the full-shape galaxy clustering measurements, e.g. see
Refs.~\cite{Ballardini:2022wzu,Mergulhao:2023ukp,Calderon:2025xod},
thus it will be promising to revisit the constraints on primordial
oscillations by using a combination of the corresponding DESI
\cite{DESI:2024hhd} and Euclid \cite{EUCLID:2011zbd} data with CMB
dataset.

It should be mentioned that our constraint on primordial
oscillations is based on the $\Lambda$CDM model which is suffering
from the Hubble tension \cite{CosmoVerse:2025txj}. Though our
combined Planck+SPT+ACT dataset showed the scalar spectral index
is $n_s\approx 0.973$, when the early dark energy resolution
\cite{Karwal:2016vyq,Poulin:2018cxd} of Hubble tension is
considered $n_s$ will be possible to shift towards $n_s=1$
\cite{Ye:2020btb,Ye:2021nej,Jiang:2022uyg,Smith:2022hwi,Jiang:2022qlj,Jiang:2023bsz,Wang:2024dka,Wang:2024tjd,Poulin:2025nfb},
see also \cite{DiValentino:2018zjj,Giare:2022rvg}. Thus
re-exploring the constraints on primordial oscillations in light
of combined Planck+SPT+ACT dataset and the potential resolutions
of Hubble tension will be also significant.

\begin{acknowledgments}
This work is supported by NSFC, No.12475064, 12075246, National Key Research and Development Program of China, No. 2021YFC2203004, and the Fundamental Research Funds for the Central Universities.
We acknowledge the use of high performance computing services provided by the International Centre for Theoretical Physics Asia-Pacific cluster and the Tianhe-2 supercomputer.
\end{acknowledgments}

\bibliography{ref}


\clearpage

\widetext
\begin{center}
\textbf{\large Supplemental Material for ``Tightening constraints on primordial oscillations with latest ACT and SPT data"}
\end{center}
\setcounter{equation}{0}
\setcounter{figure}{0}
\setcounter{table}{0}
\setcounter{page}{1}
\makeatletter
\renewcommand{\theequation}{S\arabic{equation}}
\renewcommand{\thefigure}{S\arabic{figure}}

\section{Full results of MCMC and best-fit parameters}

The full results of our MCMC analysis, including the best-fit
parameters and $\chi^2$ values for models with primordial
oscillations, are presented in the Tables~\ref{tab:pl},
\ref{tab:log} and \ref{tab:lin}. We can see that the inclusion of
primordial oscillations has negligible impact on the results of
other cosmological parameters.

\begin{table}[htbp]
    \centering
    \begin{tabular}{|l|c|c|c|}
        \hline
        Parameter & Planck & SPT+ACT & Planck+SPT+ACT \\
        \hline
        $10^9A_\mathrm{s}$  & $2.115(2.113)\pm 0.029$ & $2.142(2.163)\pm 0.024$ & $2.141(2.142)^{+0.025}_{-0.029}$ \\
        $n_\mathrm{s}$      & $0.9702(0.9690)\pm 0.0033$       & $0.9739(0.9751)\pm 0.0057$       & $0.9734(0.9732)\pm 0.0030$ \\
        $H_0$                & $68.31(68.25)\pm 0.29$          & $68.09(68.19)\pm 0.26$          & $68.11(68.14)\pm 0.25$ \\
        $100\Omega_\mathrm{b} h^2$        & $2.251(2.248) \pm 0.012$         & $2.247(2.247) \pm 0.013$         & $2.2455(2.2455) \pm 0.0090$ \\
        $\Omega_\mathrm{c} h^2$ & $0.11783(0.011792)\pm 0.00064$  & $0.11819(0.11785)\pm 0.00066$     & $0.11801(0.11792)\pm 0.00061$ \\
        $\tau_\mathrm{reio}$& $0.0601(0.0558)\pm 0.0070$ & $0.0646(0.0691)\pm 0.0059$    & $0.0650(0.0641)^{+0.0066}_{-0.0075}$ \\
        \hline
        $\chi^2$            & $2231.40$                         & $1775.00$                       & $2420.51$ \\
        \hline
    \end{tabular}
    \caption{The mean (best-fit) $\pm 1\sigma$ errors of relevant parameters and $\chi^2$ values for models with the standard PPS. }
    \label{tab:pl}
\end{table}

\begin{table}[htbp]
    \centering
    \begin{tabular}{|l|c|c|c|}
        \hline
        Parameter            & Planck                & SPT+ACT               & Planck+SPT+ACT        \\
        \hline
        $A_\mathrm{log}$     & $< 0.0387(0.0196)$            & $< 0.0335(0.0153)$            & $< 0.0286(0.0145)$            \\
        $\omega_\mathrm{log}$& unconstrained$(31.46)$         & unconstrained$(36.23)$         & unconstrained $(31.44)$        \\
        $\phi_\mathrm{log}$  & unconstrained$(4.39)$         & unconstrained$(6.13)$         & unconstrained$(4.23)$         \\
        \hline
        $10^9A_\mathrm{s}$   & $2.113(2.114) \pm 0.029$     & $2.144(2.162)\pm 0.025$     & $2.141(2.140) \pm 0.027$     \\
        $n_\mathrm{s}$       & $0.9703(0.9716) \pm 0.0033$   & $0.9733(0.9728)\pm 0.0060$ & $0.9733(0.9727) \pm 0.0029$ \\
        $H_0$                & $68.31(68.44) \pm 0.29$      & $68.11(68.24) \pm 0.26$      & $68.11(68.16) \pm 0.25$      \\
        $100\Omega_\mathrm{b} h^2$ & $2.248(2.259) \pm 0.013$ & $2.249(0.2257) \pm 0.013$    & $2.2462(2.2498) \pm 0.0097$   \\
        $\Omega_\mathrm{c} h^2$ & $0.11783(0.11766) \pm 0.00063$ & $0.11817(0.11805) \pm 0.00065$ & $0.11803(0.11798) \pm 0.00061$ \\
        $\tau_\mathrm{reio}$ & $0.0597(0.0579) \pm 0.0069$   & $0.0646(0.0686)\pm 0.0060$   & $0.0648(0.634) \pm 0.0070$   \\
        \hline
        $\chi^2$           & $2423.140$             & $1768.89$            & $2412.18$            \\
        \hline
    \end{tabular}
    \caption{The mean (best-fit) $\pm 1\sigma$ errors of relevant parameters and $\chi^2$ values for models with logarithmic oscillations. For upper limits, we report the one-side $95\%$ confidence level ($2\sigma$). }
    \label{tab:log}
\end{table}

\begin{table}[htbp]
    \centering
    \begin{tabular}{|l|c|c|c|}
        \hline
        Parameter              & Planck                & SPT+ACT               & Planck+SPT+ACT        \\
        \hline
        $A_\mathrm{lin}$       & $< 0.0357(0.0136)$            & $< 0.0433(0.0086)$            & $< 0.0267(0.0229)$            \\
        $\omega_\mathrm{lin}$  & unconstrained$(10.92)$         & $56(5.41)^{+40}_{-50}$      & unconstrained$(75.23)$         \\
        $\phi_\mathrm{lin}$    & unconstrained$(4.01)$         & unconstrained$(6.20)$         & unconstrained$(6.28)$         \\
        \hline
        $10^9A_\mathrm{s}$     & $2.113(2.104)^{+0.027}_{-0.031}$ & $2.142(2.159)\pm 0.024$   & $2.141(2.126)^{+0.024}_{-0.028}$ \\
        $n_\mathrm{s}$         & $0.9701(0.9706)\pm 0.0033$    & $0.9742(0.9766)\pm 0.0057$    & $0.9734(0.9728)\pm 0.0030$    \\
        $H_0$                  & $68.30(68.26)\pm 0.29$       & $68.10(68.22)\pm 0.26$       & $68.10(68.11)\pm 0.25$       \\
        $100\Omega_\mathrm{b} h^2$ & $2.250(2.243) \pm 0.013$  & $2.247(2.247) \pm 0.012$     & $2.2457(2.246) \pm 0.0094$   \\
        $\Omega_\mathrm{c} h^2$ & $0.11784(0.11780)\pm 0.00063$ & $0.11816(0.11785)\pm 0.00067$  & $0.11803(0.11805)\pm 0.00061$  \\
        $\tau_\mathrm{reio}$   & $0.0594(0.0559)^{+0.0064}_{-0.0074}$ & $0.0645(0.0693)\pm 0.0059$ & $0.0649(0.0605)^{+0.0062}_{-0.0073}$ \\
        \hline
        $\chi^2$            & $2420.42$             & $1767.84$            & $2414.09$            \\
        \hline
    \end{tabular}
    \caption{The mean (best-fit) $\pm 1\sigma$ errors of relevant parameters and $\chi^2$ values for models with linear oscillations. For upper limits, we report the one-side $95\%$ confidence level ($2\sigma$). }
    \label{tab:lin}
\end{table}

\end{document}